\documentstyle[twocolumn,aps,pra,epsfig]{revtex}
\def\be{\begin{equation}}
\def\e#1{\label{#1}\end{equation}}
\def\bea{\begin{eqnarray}}
\def\ea#1{\label{#1}\end{eqnarray}}

\def\ee{\end{equation}}
\def\eea{\end{eqnarray}}
\def\bem#1{\begin{mathletters}\label{#1}}
\def\eml{\end{mathletters}}

\begin{document}

\title{Self-Binding Transition in Bose Condensates with Laser-Induced 
``Gravitation''}
\author{S. Giovanazzi \and D. O'Dell \and G. Kurizki }
\address{
Department of Chemical Physics, The Weizmann Institute of Science,
76100 Rehovot, Israel}
\date{\today}
\maketitle

\begin{abstract}
In our recent publication 
(D. O'Dell, {\it et al}, Phys.\ Rev.\ Lett.\ \textbf{84}, 5687 (2000)) 
we proposed a scheme for electromagnetically generating 
a self-bound Bose-Einstein condensate with 1/r attractive interactions: 
the analog of a Bose star.
Here we focus upon the conditions neccessary
to observe the transition from external trapping to 
self-binding. 
This transition becomes manifest in a sharp reduction of the condensate
radius and its dependence on the laser intensity rather 
that the trap potential.
\end{abstract}

\section{Introduction}
\label{sec.1}

We have recently proposed \cite{odell2000} a scheme for inducing
a $1/r$ gravitational-like attractive interatomic potential in an atomic 
Bose-Einstein condensate (BEC) \cite{bec}
contained in the near-zone volume
of intersecting triads of orthogonal laser beams.
For sufficiently strong self-``gravitation'' the BEC becomes self-bound.
In this unique, novel regime the $1/r$ attraction balances the 
outward pressure due to the zero point kinetic energy 
and the short range s-wave scattering.
Here we focus upon the  transition from external trapping to 
self-binding.
This transition becomes manifest in a sharp reduction of the condensate
radius and its dependence on the laser intensity rather 
than the trap potential.
We analyze the conditions for the observability
of the self-binding transition: the threshold laser intensity
(Sec.\ \ref{sec.threshold}),
the bounds on the number of atoms 
imposed by the near-zone condition (Sec.\ \ref{sec.number}),
as well as the loss rates (Sec.\ \ref{sec.loss}).
Sec.\ \ref{sec.summary} summarizes the findings.

\section{Self-Binding Threshold Intensity}
\label{sec.threshold}

\subsection{Threshold condition}
\label{sec.2a}

We need to find a situation where the mean-field self-``gravitation''
energy associated 
with the near-zone laser-induced attractive $1/r$
potential can become (at least) comparable with
the short-range s-wave scattering energy.
To this end, we examine the mean-field solution for a condensate of atoms 
interacting via
Thirunamachandran's isotropic two-atom potential \cite{thiru80},
obtained by directional averaging of the laser-induced dipole-dipole potential.
This potential has the form 
\bea
U_{\mathrm iso}(\tilde{r}) 
= &-& \frac{15 \pi u}{11 \lambda_{\mathrm L}}
      \bigg( \frac{\sin(4\pi \tilde{r})}{(2\pi \tilde{r})^2} 
      + 2 \frac{\cos(4\pi \tilde{r})}{(2\pi \tilde{r})^3} -
        \nonumber \\
 &-&  5 \frac{\sin(4\pi \tilde{r})}{(2\pi \tilde{r})^4} 
- 6 \frac{\cos(4\pi \tilde{r})}{(2\pi \tilde{r})^5} +
        3 \frac{\sin(4\pi \tilde{r})}{(2\pi \tilde{r})^6} \bigg)
\label{averaged}
\eea
where $\tilde{r}=r/\lambda_{\mathrm L}$, is normalized to 
the laser wavelength $\lambda_{\mathrm L}$, and 
\be
u = (11 \pi/15) (I \alpha^2 \,/\, c \epsilon_0^2 
\lambda_{\mathrm L}^2) \;,
\label{udef}
\ee
$I$ being the sum of the intensities of all the lasers, and $\alpha$ the
atomic polarizability.
The potential begins to oscillate
(i.e.\ becomes alternatingly repulsive and attractive) at distances beyond
 $\sim 0.36 \lambda_{\mathrm L}$. 
However, this potential can support a self-bound condensate
with a larger radius as shown below.

We use the mean-field approximation (MFA), as embodied
in the following generalized Gross-Pitaevskii equation \cite{odell2000},
to calculate the ground-state order parameter $\Psi({\bf R})$ of a
BEC subject to a laser-induced interatomic interaction
\be
\mu \Psi ({\bf R}) =
\left[
- \frac{\hbar^{2}}{2m} \nabla^{2}
+ V_{\mathrm{ext}}({\bf R})
+ V_{\mathrm{sc}}({\bf R})
\right] \Psi({\bf R})  \label{21}
\end{equation}
where $m$ is the atomic mass, 
$V_{\mathrm{ext}}(R) = m\omega_{0}^2 R^2/2$
is an isotropic  external trap potential
(which will be considered negligible---see below), 
and
 $V_{\mathrm{sc}}({\bf R})$ is the self-consistent potential
\begin{equation}
V_{\mathrm{sc}}({\bf R})=
 g \rho({\bf R})
+ \int d^3 R^{\prime}\;
U_{\mathrm iso}( {\bf R}^{\prime}- {\bf R} )
\;\rho({\bf R^{\prime}})  \;
\label{22}
\end{equation}
where $\rho({\bf R})= \Psi^{2}({\bf R}) $ is the density  
and  $g=4 \pi a \hbar^2/m $, $a$ being the s-wave scattering
length.

In cold dilute atomic BECs with short-range s-wave scattering, the
validity of the MFA
(i.e.\ the Gross-Pitaevskii equation \cite{pitaevskii61}), 
is well established
providing  $\rho a^{3} \ll 1$.
However,
the MFA is also valid for the long range repulsive coulomb 
potential, $+u/r$, 
provided many atoms lie
within an interaction sphere with a Bohr-type radius,
$a_{\ast}=h^2/mu$, so that $\rho a_{\ast}^3 \gg 1$ \cite{foldy61}.
This condition means that the potential must be weak. Remarkably,
self-gravitating BECs simultaneously satisfy \emph{both} of these 
MFA validity conditions
as can be readily verified using the ensuing
expressions.

The condensate radius can be studied 
using the variational wavefunction  
$\Psi_{w}(R)=$ $\sqrt{N}$ $\exp (-R^2/2 w^2 \lambda_{\mathrm L}^2 )/$
$(\pi w^{2} \lambda_{\mathrm L}^{2})^{{3\over4}}$, 
where $w$ is a dimensionless variational parameter giving the width 
of the condensate.  
The variational solution in the limit of negligible kinetic energy
(Thomas-Fermi limit)
yields a self-bound condensate, 
i.e.\ \emph {finite}  $w$ (see Fig.\ 1 and Fig.\ 2 below), if  the laser 
intensity exceeds the following threshold value (in S.I. units)
\be
I_{0} = \frac {48\pi}{7}
\frac{\hbar^2  c \epsilon_0^2}{m \alpha^2} a \;.
\label{i0}
\ee
Here $I_{0}$ is the total intensity supplied by all the laser beams:
for a triad each laser should have 1/3 of the above value
and for the 6 triad configuration \cite{odell2000}  
12 of the lasers should have 1/15, and the
remaining 6 should have 1/30, of the above value. 
The threshold $I_0$ signifies the equality of the gravitational-like
potential and the s-wave scattering potential.

With an intensity 1.5 times the threshold value (Eq. (\ref{i0})) 
(arrow in Fig 2)
the expectation value of the rms condensate radius 
$R_{rms}=\sqrt{ \langle R^{2} \rangle}$
is a fraction of the laser wavelength $\lambda_{\mathrm L}$
($ R_{rms} \approx 0.43 \times \lambda_{\mathrm L}$).
The condensate is less and less confined 
as one approaches the threshold (\ref{i0})---see Fig.\ 2, from above.
Increasing the intensity $I$ reduces the condensate radius 
which becomes, in the  asymptotic limit, proportional to $1/\sqrt{I}$.
Thus the dependence $R_{rms}\sim (I_0/I)^{1/2}\lambda_{\mathrm L}$
is a distinct experimental signature of self-binding.


\begin{figure}[h]
\vspace{-0.5cm}
\begin{center}
\centerline{\psfig{figure=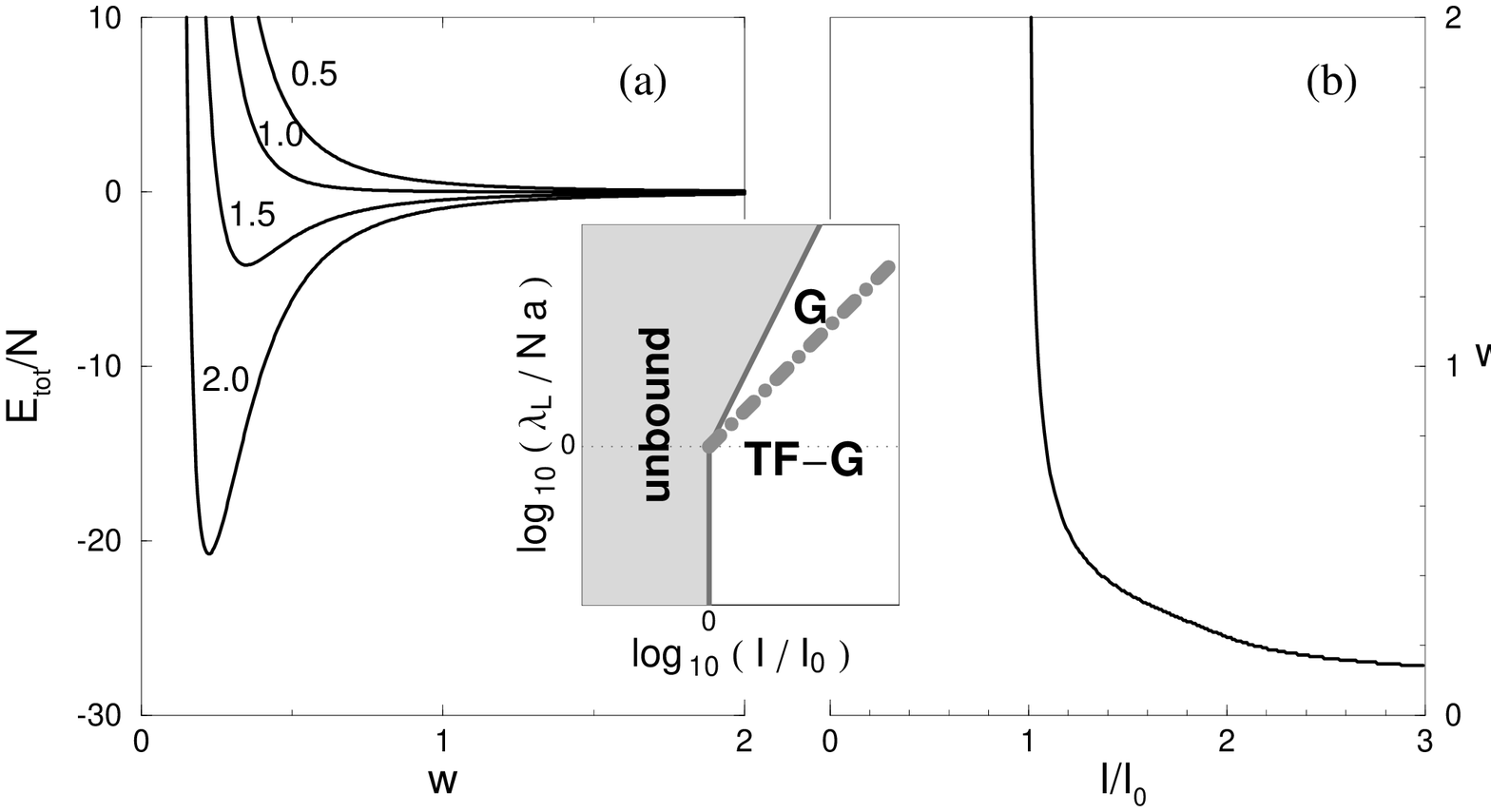,width=8.5cm}}
\end{center}
\vspace{-0.5cm}
\begin{caption}
{ (a)
Variational mean field energies per particles 
in the case of negligible kinetic energy
(TF-G regime) and $\lambda_{\mathrm L}/ N a \ll 1$
plotted versus the trial size $w$  for different values of  $I/I_0$.
  (b)
Equilibrium value of $w$ versus  $I/I_0$ in the limit of 
negligible kinetic energy (Thomas-Fermi limit).
Only for $I/I_0>1$ 
are \emph{self-bound} variational solutions 
(having minimum at finite $w$)
observed.
  Inset -
Schematic phase portrait of the transition from
unbound to self-bound regime for negligible external trapping 
is plotted versus $\log _{10}(\lambda_{\mathrm L}/ N a)$
and $\log _{10}(I / I_0)$.
}
\end{caption}
\end{figure}

At the threshold intensity an external harmonic trap becomes negligible 
when $\rho l_0 \lambda_{\mathrm L} a \gg 1$, where $l_0^2= \hbar /m\omega_0$
and $\rho$ is the density.
As the laser intensity is increased beyond this value the trap
becomes increasingly ``irrelevant''---\emph{it is not necessary} 
to turn it off to access the TF-G regime,  
where $r^{-1}$ and s-wave scattering dominate.

The threshold $I_0$ (\ref{i0}) is evaluated neglecting the kinetic energy.
The role of kinetic energy can be discussed in terms of 
$\lambda_{\mathrm L}/ N a$
(approximately the ratio between the kinetic energy 
$N\hbar^2/m\lambda_{\mathrm L}^2$
and the scattering energy
$N^2\hbar^2 a/m\lambda_{\mathrm L}^3$), as  shown schematically in the
phase portrait in Fig.\ 1 (drawn for negligible external trapping) 
which can modify the threshold for self-binding.
The G regime, representing
the purely ``gravitational'' counterpart of the TF-G regime,  
where only ``self-gravitation'' and kinetic energy play a role \cite{odell2000}
(as in a Bose star \cite{ruffini69}) is 
accessed when
\be
\frac{\lambda_{\mathrm L}} {N a} 
\lesssim \frac{I}{I_{0}} \lesssim
\left(\frac{\lambda_{\mathrm L}} {N a}\right)^2 
\ee
that implies $ 1 \lesssim \lambda_{\mathrm L} / N a  $.


At this point the variety of choices can be mainly divided into
two categories: 
i)  
to work with long laser wavelengths
in order to contain many atoms within the near zone, at the price of
very high threshold power;
ii) 
use laser wavelengths moderately detuned from an atomic
resonance,
so as to benefit from the increased polarizability,
at the price of considerably fewer self-bound atoms.

\subsection{Long-wavelength (static polarizability) threshold}
\label{sec.threshold.b}

The threshold intensity (Eq. (\ref{i0})) is 
\emph {independent of the laser wavelength} $\lambda_{\mathrm L}$,
as long as the dynamic polarizability $\alpha(q)$ is too.
The $I_0$ threshold takes the following zero-frequency (static)
values: 
$I_{0} = 5.65 \times 10^9 \; {\mathrm{Watts/cm}}^{2}$
for sodium,
$I_{0} = 8.19   \times 10^8 \; {\mathrm{Watts/cm}}^{2}$
for rubidium.
It is sufficient to use 20 W  $\times$ 3 beams of Nd:Yag lasers focused down
to 10 $\mu$m for rubidium to exceed the threshold.
By contrast, we require multi kW CO$_2$ lasers focussed down
to 100 $\mu$m for the same purpose.
A laser beam with a gaussian profile
focussed to 10 $\lambda_{\mathrm L}$ would exert a
large inward radial dipole force on each atom, so 
non-gaussian optics giving a very flat intensity profile \cite{chen95}
over the condensate region may be required 
in the long-wavelength (static) case.
There remains the problem of random noise in the intensity profile, 
but fortunately this can only exist on scales larger than 
the wavelength and so may be overcome.

An additional option is to \emph {reduce the scattering length} $a$
(to which the threshold intensity (\ref{i0}) is proportional).
This is possible in the vicinity of (but somewhat off) a Feshbach resonance,
as demonstrated experimentally \cite{inouye9899}:
reduction of $a$, and correspondly $I_0$, by one to two orders of magnitude
would eliminate the need for non-gaussian optics
in the static polarizability case.


\subsection{Moderate-detuning threshold}
\label{sec.threshold.c}

Using a moderate detuning from an atomic resonance
one can increase the polarizability 
by many orders of magnitude compared to its zero frequency value.

In a recent experiment on superradiance 
\cite{inouye2000}
the laser was red detuned by 1.7 GHz from the 3S$_{1/2}$, 
F=1 $\rightarrow$ 3P$_{3/2}$, F = 0,1,2, transition of sodium.
With this detuning, the polarizability 
in cgs units is 
$\alpha = 3.534 \times 10^{-18} {\mathrm cm}^{3}$,
which is $\approx 1.5 \times 10^{5}$
times  the static value of the polarizability.
The threshold intensity (\ref{i0})
is then reduced by a factor $\approx 2.3\times10^{10}$
compared to the static polarizability case, becoming
$I_{0} \approx 262  \;\; {\mathrm{mW/cm}}^{2}$
for sodium, which is close to the values used in Ref.\ 
\cite{inouye2000}.

With this value of threshold intensity the gradient forces 
can be negligible if the focal spots of 
the lasers are much wider than $\lambda_{\mathrm L}$.


\subsection{Moderate-detuning saturation and repulsion}
\label{sec.threshold.d}

The potential (\ref{averaged}) is the result of
a 4th order, two-atom, QED 
process \cite{thiru80}, valid when the laser
is \emph{far detuned} from any atomic transitions. 
This means that the initial absorption of a laser photon
and the subsequent intermediate steps are virtual processes
(which are most significant in the near-zone),
followed by photon emission back into
the original laser mode.
A different process can take place when
the laser is on-resonance.
Genuine absorption of a laser photon by a single atom
(measured by the saturation), followed by spontaneous
emission of this real photon is a process that
radiates energy. 
If another atom absorbs this radiation then in the far-zone it feels
a repulsive Coulomb-like
force $F_{\mathrm repuls} = K/r^2$ \cite{walker90},
which has been recently measured in rubidium molasses
\cite{pruvoust2000}. For moderate detuning, can this force
counteract our attractive gravitation-like force
$F_{\mathrm grav} = -u/r^2$?

For detuning $\delta$ 
much larger than both the Rabi frequency $\Omega$ and 
the linewhidth $\gamma$ of the resonance, the 
saturation parameter 
$s =  I d^2/(\epsilon_0 c \hbar^2 \delta^2)$
\cite{cctvol2},
where $d$ is the dipole matrix element,
becomes \emph {independent of the 
detuning}  
when calculated at the threshold intensity (\ref{i0})
\be
s(I=I_0) = \frac{48\pi}{7}\frac{a \epsilon_0  \hbar^2}{m d^2}.
\label{eq:saturation}
\ee

 It is then found that \cite{pruvoust2000}
$K \approx \sigma_0^2 I_s \Omega^4/(16 c \delta^2)$,
where $\sigma_0$ is the resonant absorption cross section and $I_s$ is the 
corresponding saturation intensity.
On comparing this expression with $u$ (Eq.\ (\ref{udef})),
we find that, in terms of the saturation parameter $s$,
\be
K \approx s u \;.
\ee
For the sodium transition
and 1.7 GHz detuning referred above, Eq.\ (\ref{eq:saturation}) yields 
very small value $s \approx 0.0003$.
This implies that under the moderate-detuning conditions discussed 
above,
the repulsive force has a \emph {negligible} effect on self-binding.

\section{Number of self-bound atoms}
\label{sec.number}

A key experimental restriction on self-binding
is that the atoms should be in the near-zone to feel the $1/r$ 
potential: a condensate smaller than the laser wavelength
 limits the number of atoms involved. 
Let us  assume we have the maximum density of
some $10^{15}$ atoms/cm$^3$. 
Using the gaussian wavefunction 
one can have of the order of $10^6$  or $10^3$
atoms in the condensate irradiated by a CO$_2$ laser or
Nd:Yag laser, respectively (see Fig. 2).

The price of moderately detuned  wavelengths 
($\approx .589 \mu$m for sodium) is the small number of atoms involved. 
With an intensity $I\approx 1.5 I_0$
the atom cloud 
contains $\approx 40$ atoms  as the  peak
density ranges from  $10^{15}$ to  $10^{16}$ atoms/cm$^3$.
Although this number is small, it is 
\emph {sufficient to demonstrate the self-binding effect}.

For given values of $I$, $\alpha$, $a$, and  $m$, 
we are either in the G regime or the TF-G regime,
depending on whether the number of atoms $N$ is 
smaller or larger than the number \cite{odell2000}
$N_{\mathrm border} \approx \sqrt{3 \pi \hbar^2/(2 m u a)}$
which corresponds to the line separating the two regions in the
inset of Fig. 1.

It so happens that 40, the lower estimate of the number of self-bound
sodium atoms obtainable in the moderate-detuning regime, 
is very close to $N_{\mathrm border}$.
This is an interesting region, because both
the kinetic energy and the s-wave scattering 
are significant and together with the $r^{-1}$ attraction 
determine the condensate properties.

\vspace{0.5cm}
\begin{figure}[h]
\vspace{-1cm}
\begin{center}
\centerline{\psfig{figure=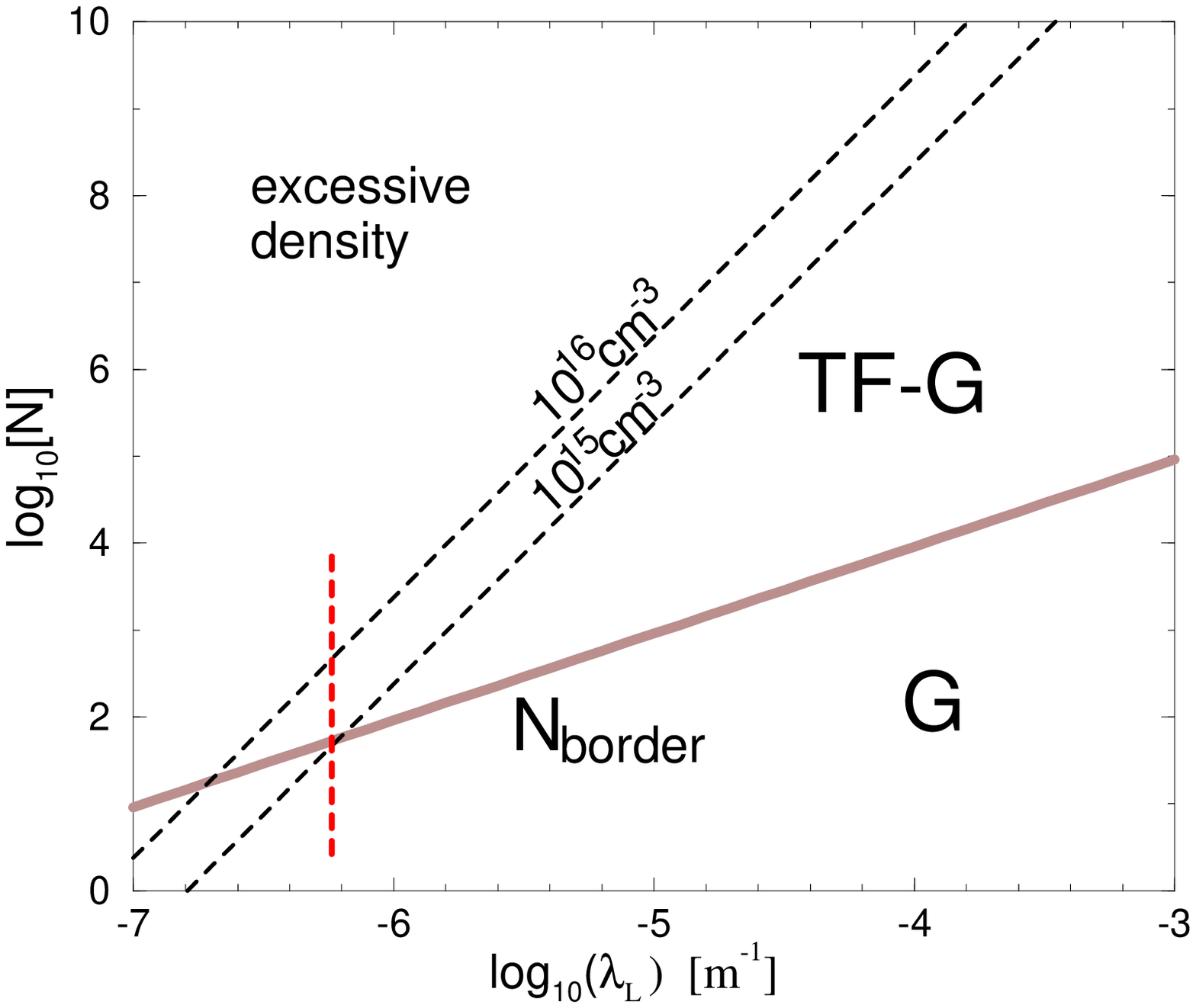,height=6.5cm}}
\end{center}
\vspace{-0.1cm}
\begin{caption}
{Range of numbers $N$ of Na condensate atoms as a function of 
$\lambda_{\mathrm L}$
that are compatible  with a TF-G or G solution. 
The  density is $10^{15}$--$10^{16}$ atoms/cm$^3$
and the intensity is $1.5$ times the threshold intensity (\ref{i0}).
The region above $10^{16}$ cm$^{-3}$ corresponds to excessive density.
The vertical long-dashed line corresponds to the moderate-detuning
choice discussed for Na.}
\end{caption}
\end{figure}

\section{Loss rates}
\label{sec.loss}

\subsection{Spontaneous Rayleigh losses}
\label{sec.3a}

The single-atom Rayleigh scattering rate  $\Gamma_{\mathrm ray}$
leads to depletion of the condensate. The probability amplitude
for inelastic scattering from the ground state $|0 \rangle$ of the
near-zone condensate to any excited state $|n \rangle$ due to an external
field with wavevector
$\mathbf{q}$ is proportional to $ \sqrt{N} \sum_{n \neq 0} \langle n |
({\mathbf q} \cdot {\mathbf r}) | 0 \rangle$. 
Hence, for sample sizes less than a wavelength we expect the spontaneous
Rayleigh scattering rate to be reduced by a factor 
at least as small as $(q\,R_{rms})^2$, analogously to the
Lamb-Dicke effect \cite{dicke53}.
The lifetime of the condensate, when
determined from spontaneous Rayleigh scattering alone, is
estimated to be 
\be
\tau_{\mathrm ray} \ge 
\left(
\Gamma_{\mathrm ray} \left(q\,R_{rms} \right)^{2}
\right)^{-1}\;.
\ee

Since $\Gamma_{\mathrm ray}=
I q^{3} \alpha^{2}/(3 h \epsilon_{0}^{2}c)$ \cite{thiru80}, 
it can be expressed in terms of the
electromagnetically induced energy $U(r)=-u/r$ of a \emph{single}
pair of atoms separated by a distance
equal to the wavelength 
\be
 \Gamma_{\mathrm ray} = \left( \frac{20 \pi}{ 11} \right) 
\frac{u}{\hbar \lambda_{\mathrm L}}
\label{gamma}
\ee
where $u$ is defined in Eq.\ (\ref{udef}). Using this relation, we 
can compare the upper bound on the condensate lifetime set by 
Rayleigh scattering  with the time scale 
of the dynamics, the requirement being that the system exists 
long enough to equilibriate.
In the TF-G and G (self-bound) regions a characteristic 
time scale for the dynamics
is provided by the following ``plasma'' frequency 
\be
\omega_{\mathrm p}^2 = \frac{4 \pi u \rho_{\mathrm peak}}{m} 
\ee
where $\rho_{\mathrm peak}$ is the peak density.
We can express $\omega_{\mathrm p}$ in terms of 
the recoil energy 
$E_{\mathrm R}=\hbar^2q^2/2m$ ($q$ being the mean laser wavelength) 
 and the Rayleigh scattering rate $\Gamma_{\mathrm ray}$
using Eq.\ (\ref{gamma}) 
\be
\omega_{\mathrm p} \approx 0.25\; 
\frac{\hbar \Gamma_{\mathrm ray}^2}{E_{\mathrm R}}
\;N^2 f^{-3/2} 
\label{eq:explicit-plasma-freq}
\ee
where the factor 
\be
f = \frac 12 +\sqrt{\frac 14 + \frac {N^2}{N_{\mathrm border}^2}}
\ee
is asymptotically equal to 1 in the G region 
and $N/N_{\mathrm border}$ in the TF-G region.
It follows from (\ref{eq:explicit-plasma-freq}) that the 
characteristic oscillation 
frequency $\omega_{\mathrm p}$ can be much bigger 
than $\Gamma_{\mathrm ray}$, 
by a factor proportional to $N^2$ or $N^{1/2}$ in the G or TF-G region,
 respectively.
Thus the lifetime can be considerably
longer than the characteristic time scale of the  dynamics.

Even for the small number of 40 sodium atoms in the self-bound
moderate-detuning  regime
($I=1.5 \times I_{0}$, $\delta = 1.7$ GHz), for which 
the recoil energy is $E_{\mathrm R}/\hbar=1.57\times 10^5$  s$^{-1}$ 
and
$\Gamma_{\mathrm ray} = 1.58 \times 10^4$ s$^{-1}$, we find   
$\omega_{\mathrm p} \approx 20 \times \Gamma_{\mathrm ray}$.
This implies that several oscillation periods of the self-bound
condensate can occur within the Rayleigh lifetime.

\subsection{Intereference losses} 
\label{sec.3b}

We revisit the expressions for the loss rate $\Gamma_{\mathrm interf}$
due to multi-beam interference as obtained in  \cite{odell2000}.
We can express $\Gamma_{\mathrm interf}$ in terms of 
the recoil energy $E_{\mathrm R}$ and Rayleigh scattering 
rate $\Gamma_{\mathrm ray}$ as in Sec. \ref{sec.3a} 
\be
\Gamma_{\mathrm interf} \approx 0.05 \; 
\left( \frac{\hbar\Gamma_{\mathrm ray}N}{E_{\mathrm R}}\right)^4
\sqrt{\frac{\hbar\Omega}{E_{\mathrm R}}}\;\Gamma_{\mathrm ray}\;f^{-3}
\ee
where $\Omega$ is the relative detuning of beams in the triad.
In the example given in Sec.\ \ref{sec.3a} above, 
$\Gamma_{\mathrm interf}$ turns out to be 
few times bigger than  
$\Gamma_{\mathrm ray}$,
when $\Omega$ is chosen to be
of the order of 
$\omega_{\mathrm p}$.

\section{Conclusions}
\label{sec.summary}

Our main conclusion is that at least the TF-G self-bound
region is experimentally accessible, although such an experiment 
would be challenging. 
Moderate detuning is preferable to the longer wavelength case 
due to the huge enhancement in the polarizability,
but it allows the self-binding of few (less that 100) atoms. 
If the scattering length were reduced via a Feshbach resonance then this would 
further facilitate the self-trapping of many more atoms using near-infrared 
lasers.

This work has been supported by the German-Israeli Foundation (GIF).

%

\end{document}